\documentclass[doublecol,figures]{epl2}

\title{Motility of small nematodes in wet granular media}
\shorttitle{Motility of small nematodes in wet granular media}

\author{G. Juarez \and K. Lu \and J. Sznitman\thanks{Present address: Department of Biomedical Engineering,
Technion - Israel Institute of Technology, Haifa 32000, Israel} \and 
P. E. Arratia\thanks{E-mail: \email{parratia@seas.upenn.edu}}}

\shortauthor{G. Juarez \etal}

\institute{Department of Mechanical Engineering and Applied
Mechanics, University of Pennsylvania, Philadelphia, PA 19104, USA\\}


\abstract{The motility of the worm nematode \textit{Caenorhabditis elegans} 
is investigated in shallow, wet granular media as a function of particle size dispersity 
and area density ($\phi$). Surprisingly, we find that the nematode's propulsion speed is
enhanced by the presence of particles in a fluid and is nearly independent of area density.
The undulation speed, often used to differentiate locomotion gaits, is significantly 
affected by the bulk material properties of wet mono- and polydisperse granular media
for $\phi \geq 0.55$. This difference is characterized by a change in the nematode's waveform 
from swimming to crawling in dense polydisperse media \textit{only}. This change highlights the 
organism's adaptability to subtle differences in local structure and response between 
monodisperse and polydisperse media.}

\begin{document}

\maketitle

\section{Introduction}

Many live organisms evolve within complex fluidic environments such
as intestinal fluid, human mucus, and soil~\cite{fauci, maladen,
vogel}. Such organisms often develop unique methods of locomotion
that allow them to exploit the material properties of the media
\cite{shimada, maladen, mazo}. For example, in the case of freely
swimming spermatozoa in a Newtonian fluid, the flagellum shows a regular
sinusoidal beating pattern in both space and time. However, once the 
organism encounters a complex fluid, like semen,
this regular beating pattern is replaced by
high-amplitude, asymmetric bending of the flagellum~\cite{ho}. This
`hyper-activated' sperm is clearly affected by its fluidic
environment, which in turn, can affect human fertility~\cite{fauci}.

There has been a renewed interest in swimming at low
Reynolds numbers~\cite{teran, lauga1, dreyfus, fauci, polin, cohen}, where
live organisms must break time-reversal symmetry~\cite{purcell} in
order to achieve net motion. The majority of previous work on
locomotion, however, has been restricted to Newtonian liquids, and
few investigations are available on the motility of live
organisms in complex fluids such as granular media \cite{maladen, mazo, shimada, jung2}.
Recent investigations \cite{shimada, li}
have indicated that in granular media, an organism needs to tune
its swimming amplitude and frequency in order to fluidize the
granular bed to minimize resistance while keeping the bed rigid
enough that it can support applied loads. For millimeter-sized undulatory
swimmers, such as the worm nematode \textit{Caenorhabditis elegans} 
(\emph{C. elegans}), experiments show that the nematode swims
more efficiently in a monodisperse particulate system of packing 
fraction $\phi=0.6$ than in a fluid without particles \cite{jung2}. 
This observed swimming enhancement was found to be independent 
of the surrounding particle size.

Despite recent efforts, many important questions remain. For instance, 
it is known that certain properties of granular media, including 
packing fraction and particle size dispersity, can significantly affect 
the response of the bulk material to external perturbations due to the
structure of force chains. For example, the response of static quasi-2D 
piles to an external force behaves more rigid with increasing $\phi$, and for 
disordered systems the response is best described by elastoplastic models
\cite{luding, geng, muthuswamy}. It is yet to be seen how the difference 
in information propagation between ordered and disordered granular
media can affect the motility of live organisms.

Here, we investigate the motility behavior of the nematode
\textit{C. elegans} in a buffer solution~\cite{brenner} of viscosity
and density similar to water. The solution contains either monodisperse or
polydisperse particles of varying area density. Such a fluid can 
exhibit both solid- and fluid-like behavior depending on packing fraction
$\phi$ (referred to here as particle area density) \cite{larson} and is 
believed to mimic more accurately the soil-like environments where 
the nematode is originally found. \textit{C. elegans} is studied
extensively and serves as a model organism, and are chosen for this
study because they are known to move in different fluidic environments \cite{korta,
sznitman3, berri, park}. They are typically 1 mm in length, 0.07 mm in diameter,
and are equipped with 95 muscle cells aligning their dorsal and
ventral sides \cite{white}. Motility is achieved from a `thrashing'
motion against the surrounding medium \cite{pierce},
and typically occurs at a low Reynolds number, 
Re $\sim \mathcal{O}(10^{-1})$ \cite{Sznitman2}. This makes \textit{C. elegans} 
an attractive candidate for motility studies. 

\section{Experimental methods}

\begin{figure}
\begin{center}
\includegraphics[width=8.4cm]{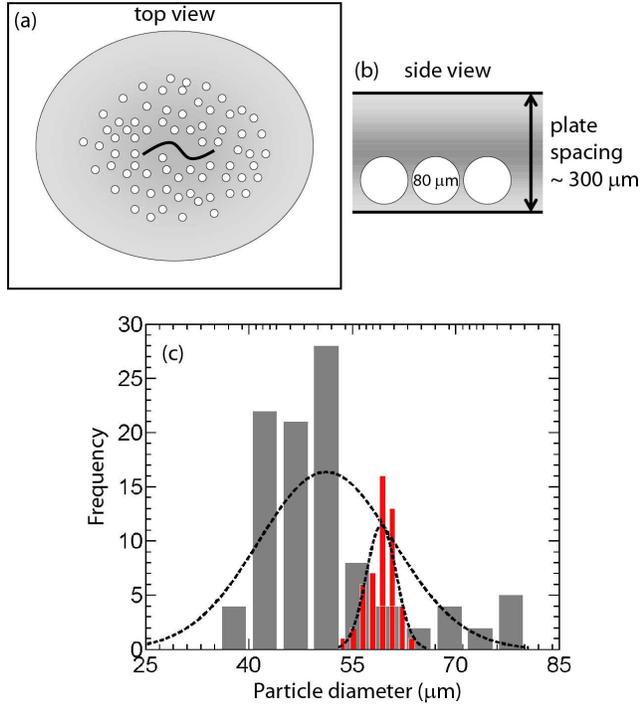}
\end{center}
\caption{(Color online) (a) Top view schematic of nematodes swimming in
buffer solution ($24 \times 40$ mm) and a quasi-2D granular layer ($12 \times 12$ mm). 
(b) Side view schematic of the plate spacing used
to confine the nematodes, buffer solution, and granular monolayer. The largest
particles used were just under 100 $\mu$m while the plate spacing was 
controlled by the volume of buffer solution and was approximately 300 $\mu$m
(c) Plot of the size distribution for monodisperse particles  
(out of 50 measurements) is shown in red, and for polydisperse particles
(out of 100 measurements) is shown in gray.}
\label{fig:sizehist}
\end{figure}

Experiments are performed in a shallow two-dimensional particle bed
in order to minimize out-of-plane undulations. The granular mono-layer 
is created by placing the nematode, buffer solution, and particles
between two glass cover-slips ($170 \ \mu$m thick), as shown in 
fig.~\ref{fig:sizehist}(a,b). The plate spacing is set by the contact angle 
between the liquid, the glass cover-slip, and the ambient air
as well as the volume of the buffer solution. A constant volume of
400 $\mu$l of buffer solution is used for each experiment to produce a plate spacing of
approximately $300 \ \mu$m, which is optically measured. The
particles then settle on the bottom plate creating a randomly packed granular 
layer in which the nematode moves free of confinement effects from the top
cover-slip, as shown in fig.~\ref{fig:sizehist}(b). Note that the total bath size
($24\times40\times0.3$ mm) is much larger than the granular bed ($12 \times 12 \times 0.08$ mm) 
and the nematode. Monodisperse samples consist of glass beads 
with diameter $d_m = 60 \pm 3 \ \mu$m (Mo-Sci Specialty Products) while
polydisperse samples consist of glass beads with diameter 
$d_p = 52 \pm 10 \ \mu$m (Potters Industries Inc.). A histogram 
with the measured size distributions for both samples is shown in fig.~\ref{fig:sizehist}(c).

Two main types of experiments are performed, body tracking methods
and particle tracking velocimetry.
The motility behavior of \textit{C. elegans} is characterized using 
body posture tracking methods \cite{sznitman}. 
The effects of the nematode's motion on the surrounding particles
are investigated using particle tracking methods \cite{crocker}.
Both \textit{C. elegans} and particles are imaged using an
epi-fluorescent inverted microscope at $5\times$ magnification (Axio
Observer Z1, Carl Zeiss Inc.) with an acquisition rate of 125 frames
per second (Fastcam SA1.1, Photron USA Inc.). The \textit{C.
elegans} strain (Genetic Stock Center) is cultured using standard
methods \cite{brenner}.

\section{Nematode kinematics}

\begin{figure}
\begin{center}
\includegraphics[width=8.4cm]{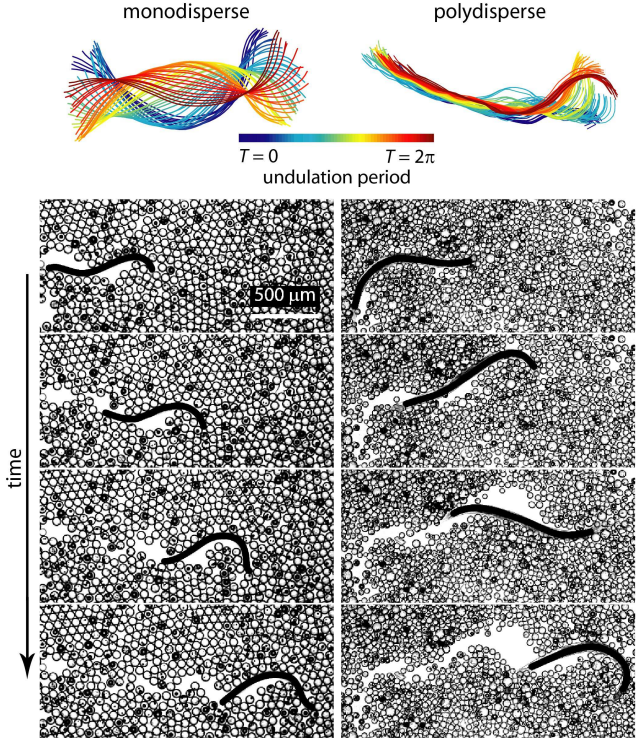}
\end{center}
\caption{(Color online) Plots of the nematode skeleton for two 
environments with particle area densities of $0.55$; (Left) 
monodisperse granular media (Right) polydisperse granular media. 
The undulation period $T$ is color-coded such that blue and red correspond 
to $T=0$ and $T=2\pi$, respectively. Below the contours are time-sequenced 
micrographs of wild-type \textit{C. elegans} moving through (Left) 
monodisperse and (Right) polydisperse wet granular media with an area density 
of $0.55$. Images are taken at equally spaced intervals 
of $\Delta t = 1.12$ s progressing from top to bottom. Note the differences 
between the bottom two micrographs: (Right) large amount of white space, or 
void space, free of particles compared to (Left)  monodisperse media.
See supplementary online material for corresponding movies.}
\label{fig:nematode}
\end{figure}

\begin{figure}
\begin{center}
\includegraphics[width=8.1cm]{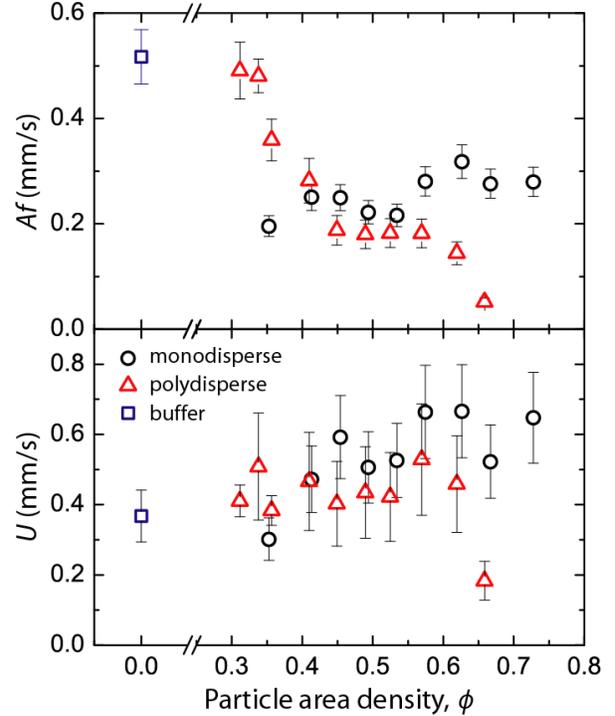}
\end{center}
\caption{(Color online) (a) Undulation speed and (b)
swimming speed as a function of local area density for \textit{C.
elegans} in wet granular media. The undulation speed decreases 
drastically in granular media for area densities above $0.3$
and remains constant in monodisperse media, but decreases again in 
polydisperse media beyond $\phi \approx 0.55$. The swimming speed 
increases slightly in granular media and remains nearly constant 
with increasing $\phi$.} 
\label{fig:velocity}
\end{figure}

The \textit{C. elegans}' kinematics in wet granular media is
analyzed by tracking the nematode body posture in both space and
time. Figure~\ref{fig:nematode} and supplementary online material
shows sample snapshots of the nematode moving in monodisperse
(left) and polydisperse (right) wet granular media and the corresponding
body postures at a particle area density of approximately 0.55. 
An important quantity that can be
obtained from body postures is the undulation speed defined as $Af$,
where $A$ and $f$ are the nematode's swimming amplitude and
frequency, respectively. The beating amplitude $A$ is defined
as half of the peak-to-peak distance swept out by the nematode head 
perpendicular to the swimming direction (fig.~\ref{fig:nematode}, top). 
The beating frequency is the inverse of the period $T$, or the time it takes 
for the head to complete one sweeping motion. The quantity $Af$ is 
often used as a relative measure of changes in motility gait \cite{pierce}.
Figure~\ref{fig:velocity}(a) shows $Af$ as a function of local area
density for both monodisperse and polydisperse media. 
We define the local area density $\phi$ of the granular layer 
as $\phi = N [\bar{d}/3L_n]^2$, where $N$ is the total number of spheres,
with average diameter $\bar{d}$ of either 62 $\mu$m for monodisperse media 
or 52 $\mu$m for polydisperse media, within an interrogation region locally
around the nematode centroid with radius of $\mathcal{R}=3L_n/4$, where
$L_n$ is a nematode body length.

Results show that the undulation speed decreases by almost one-half in wet granular
media for area density ranging from 0.3 to 0.5. The undulation
speeds are $0.52 \pm 0.05$ mm/s and $0.23 \pm 0.05$
mm/s for buffer solutions ($\phi=0$) and granular media (mono- and
polydisperse media), respectively. This decrease in $Af$ indicates
that the nematode moves with slower and/or smaller undulations. For
values of $\phi$ larger than 0.55, the undulation speed is also affected
by the media dispersity (fig.~\ref{fig:velocity}a). For the
polydisperse case, the undulation speed decreases for
$\phi\geq0.55$, whereas $Af$ remains nearly constant for the
monodisperse case under similar conditions.

The decrease in undulation speed for area densities larger than 
0.55 is accompanied by the nematode's change in motility gait.
The change in motility behavior can be seen at the top of
fig.~\ref{fig:nematode}, which shows the time-evolutions of
nematode skeletons over one beating cycle. The contour lines are
evenly spaced with intervals of 0.08 s, centered at the nematode's
center-of-mass, with the head oriented to the right. The nematode
exhibits markedly different waveforms in a buffer solution without
particles ($\phi=0$) than in the same solution containing a high
concentration of polydisperse particles ($\phi \geq 0.55$). For the
monodisperse case, the nematode exhibits a waveform identical to 
the buffer solution ($\phi=0$) for all values of $\phi$ investigated
here. The near-symmetrical waveform seen during swimming
(fig.~\ref{fig:nematode}, top-left) in a buffer solution or
monodisperse media is transformed into an apparent slithering
motion as the nematode responds to a dense polydisperse medium
(fig.~\ref{fig:nematode}, top-right). The peak-to-peak amplitudes
for the two cases are approximately $0.36$ mm for $\phi = 0$ and
$0.175$ mm for $\phi > 0.55$. This change in gait was not
observed in monodisperse media, and it is indicative of the
differences in response and material properties between the 
two granular media, discussed in more detail in the following section.

\begin{figure}
\begin{center}
\includegraphics[width=8.4cm]{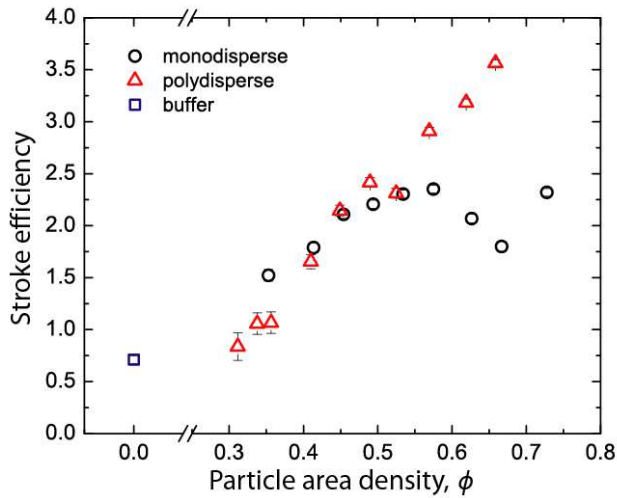}
\end{center}
\caption{(Color online) Stroke efficiency, the ratio of propulsion
speed $U$ to undulation speed $Af$, as a function of local particle 
area density for \textit{C. elegans} in wet granular media. The Stroke 
efficiency increases with area density up to $0.55$ and 
then remains constant at a value of approximately 2.25 for monodisperse 
media. The Stroke efficiency increases monotonically for polydisperse media.} 
\label{fig:Steff}
\end{figure}

Figure~\ref{fig:velocity}(b) shows the nematode's propulsion speed
$U$ as a function of $\phi$ for monodisperse and polydisperse media.
Results show that $U$ increases in wet granular media compared to swimming in a
buffer solution, and remains relatively constant, even as $\phi$
increases. The average nematode propulsion speed in buffer solution
is $0.36 \pm 0.01$ mm/s, whereas the propulsion speed in wet
granular media is $0.48 \pm 0.12$ mm/s, varying slightly with local
area density. Overall, we find that even though the undulation speed
decreased in dense wet granular media, this did not hinder the
\textit{C. elegans} ability to move. This observation can be further
quantified by computing the ratio of propulsion speed $U$ to undulation
speed $Af$ as a function of $\phi$, referred to here as the Stroke 
efficiency $\nu = U/Af$, shown in fig.~\ref{fig:Steff}. Values
of $\nu > 1$ imply that there is greater forward swimming speed 
than transverse undulation speed. We find that 
$\nu \approx 0.69$ for swimming in a fluid without particles and 
$\nu \approx 2.08$ for swimming in wet granular media. The Stroke efficiency 
increases with increasing $\phi$ implying that the nematode moves more
efficiently by generating more forward motion with less lateral motion.
Obviously, the efficiency cannot be unbounded and is expected to plateau
at some value of $\phi$ until reaching the jamming limit. Monodisperse media
appears to plateau to a value of $\nu \approx 2.25$ for area densities
greater than $0.55$. We note that the Strouhal number, 
which is the inverse of $\nu$, is typically used to characterize vortex 
shedding in oscillating flow mechanisms. It was also recently 
used to suggest that \textit{C. elegans} move more efficiently in granular media \cite{jung2}.

\section{Nematode propulsion in granular media}

\begin{figure}
\begin{center}
\includegraphics[width=8.3cm]{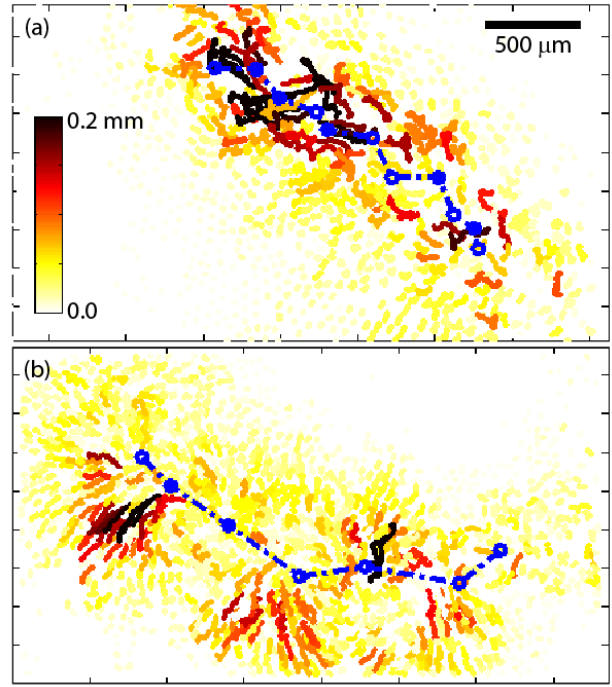}
\end{center}
\caption{(Color online) Particle pathlines induced by the nematode in
(a) monodisperse and (b) polydisperse media for $\phi \approx 0.55$.
The corresponding nematode undulation frequencies are
 $f_m \approx 1.5$ Hz and $f_p \approx 0.5$ Hz. Particles are tracked 
 for $t=3.1$ s. Nematode trajectories are depicted by the blue line with circles 
representing the nematode center of mass. The length of particle 
trajectories is color coded such that white and red correspond to 
short and long trajectories, respectively. The difference in rearrangement
to nematode undulations is evident. Monodisperse
particle trajectories are localized around the main nematode trajectory, whereas 
polydisperse particle trajectories extend outward normal to the main nematode 
trajectory. See supplementary online material for corresponding movies.} 
\label{fig:structure}
\end{figure}

\begin{figure*}
\begin{center}
\includegraphics[width=\linewidth]{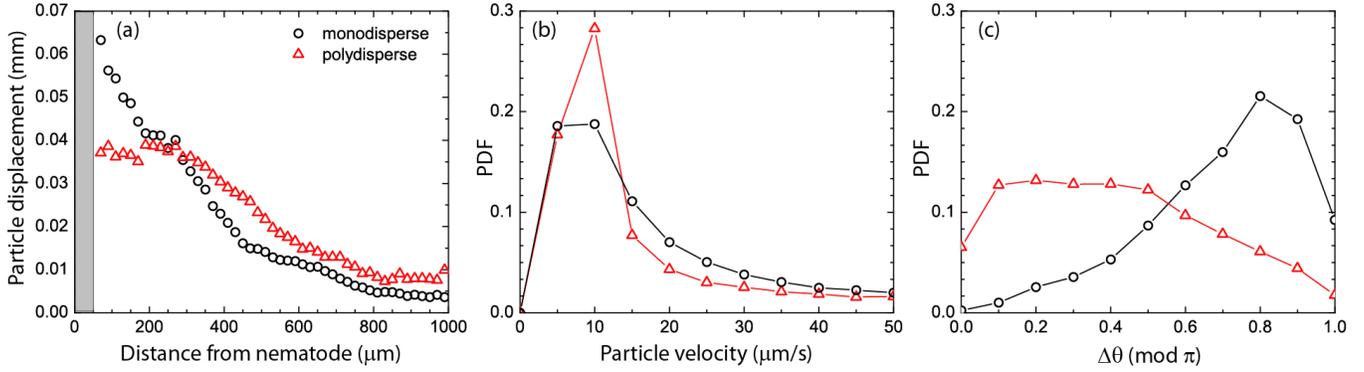}
\end{center}
\caption{(Color online) (a) Average overall particle displacement as a function of
the radial distance away from the nematode body; the gray region corresponds to the
body thickness. Particle displacement is largest near the nematode body and
decays rapidly for monodisperse media compared to the polydisperse media. (b) 
PDF of particle velocities localized around the nematode trajectory. 
While particle trajectories are longer for monodisperse media (see fig.~\ref{fig:structure}),
the probability density function (PDF) of the particle velocities are
similar for both monodisperse and polydisperse media. (c) PDF of $\Delta
\theta \ ($mod $\pi)$, the angle between the nematode trajectory and
surrounding instantaneous particle velocities. The difference in rearrangement
between the two media is evident; the majority of monodisperse particles
move nearly anti-parallel to the nematode main trajectory while polydisperse
particles move perpendicular to the nematode main trajectory.
Note that the lines are added to guide the eye.} \label{fig:trjct}
\end{figure*}

It is known that \textit{C. elegans} have distinct motility gaits,
such as swimming or crawling, depending on whether the environment is a liquid
buffer solution or a solid agar gel, respectively \cite{pierce}. That is, the
nematode is sensitive to the material properties of its surrounding fluidic
environment. Here, for $\phi < 0.3$, particles are sparse (dilute)
and the nematode behaves essentially as if it swimming in a pure buffer solution
with no particles. In addition, figs.~\ref{fig:velocity} and~\ref{fig:Steff} show 
that for area densities in the range of $0.3 < \phi < 0.55$, the nematode motility behavior
is similar for both mono- and polydisperse media which indicates that both
media have similar material properties. This is due to the lack of large scale 
structure formation in both media for this semi-dilute density range.
However, for $\phi > 0.55$, there is an obvious difference in motility behavior
that is specific to mono- or polydisperse media as noted by the wave form
(fig.~\ref{fig:nematode}, top) and undulation speed (fig.~\ref{fig:velocity}). 
This suggests that in this dense regime ($\phi>0.55$) the bulk material properties 
of polydisperse media are similar to an elastic material \cite{geng, bouchaud, goldenberg} such as 
an agar gel, while monodisperse media behave similar to a viscous 
fluid~\cite{jaeger}. 

In order to gain further insight into the difference in response of mono- and
polydisperse media to the nematodes undulations for $\phi > 0.55$, 
we tracked an average of 2,000 particles along the nematodes path. 
Figure~\ref{fig:structure} shows particle
pathlines for monodisperse (fig.~\ref{fig:structure}a) and
polydisperse (fig.~\ref{fig:structure}b) media as a nematode moves
through a $3.1$ mm by $1.8$ mm region for $\phi \approx 0.55$.
For the monodisperse case, most of the long particle trajectories
are localized around and are tangential to the nematode's body. By contrast,
for the polydisperse case, particle displacements extend further
away from and are mostly normal to the nematode's body. The average
displacement of particles nearest the nematode is $\Delta \bar{x}_m
\approx 60\ \mu$m and $\Delta \bar{x}_p \approx 40\ \mu$m for
monodisperse and polydisperse media, respectively. 
The average particle displacement decreases rapidly with radial
distance from the nematode's body (fig.~\ref{fig:trjct}a); 
this decay is expected for viscous fluids. The average particle displacement 
decays to one-half of its magnitude at 
a distance of $0.3 L_n$ for monodisperse media compared 
to a distance of $0.55 L_n$ for polydisperse media, where $L_n=1$ mm.
As expected, the distributions of particle velocities for both media 
are similar, which indicates similar energy input from nematode 
undulations (fig.~\ref{fig:trjct}b).

In general, we find that there is a noticeable difference 
in particle rearrangement behavior between mono- and polydisperse media due to
nematode undulations at comparable $\phi$. We characterize this
difference by computing the angle between the nematode trajectory
and the instantaneous velocity of the surrounding particles, $\Delta
\theta$ (mod $\pi)$, shown in fig.~\ref{fig:trjct}(c). Here, the values of
 $\Delta \theta$ (mod $\pi) = 0^\circ$ and $\Delta \theta$ (mod $\pi) = 180^\circ$
correspond to the same and opposite direction of swimming, respectively.
The data shows that for the monodisperse medium, there is a sharp 
peak at 0.8 or $\Delta \theta \approx 150^\circ$ which indicates that 
the particles move along the nematode's main trajectory. By
contrast, for the polydisperse medium, there is a broad distribution
of particle directions mostly moving normal to the nematode's main
trajectory. The sample snapshots of the nematode moving in dense granular media
(fig.~\ref{fig:nematode} and supplementary online movies) illustrates the 
difference in rearrangement behavior. For the monodisperse 
case, we observe that the local area density at the tail is only slightly
less than at the head (fig.~\ref{fig:nematode}, bottom-left) indicating 
that particles are able to easily rearrange during the perturbation imposed 
by the moving nematode. This behavior is similar to what is observed for an intruder 
dragged at constant force through an amorphous monolayer of vibrated grains 
\cite{candelier}. For the polydisperse case, however,
the overall motion of both small and large particles is mostly 
laterally from the trajectory of a nematode (fig.~\ref{fig:nematode}, bottom-right)
and particles do not readily fill the voids, leaving a dilute path behind 
the nematode. This observation provides further evidence that 
polydisperse media behaves elastoplastically, by retaining deformations due to 
nematode undulations.

Overall, we find that \textit{C. elegans'} are able to adapt and 
utilize the surrounding granular media to convert lateral undulations into 
net forward motion. This conversion leads to an increase in propulsion 
speed, which the nematode maintains even with increasing particle 
area density and degree of size dispersity. The observed enhanced propulsion 
may be explained by closely inspecting the nematode skeleton during swimming 
in buffer solutions, (fig.~\ref{fig:nematode}, top-left), and recognizing
that the motion over one beating cycle is mostly in the lateral direction. 
As $\phi$ increases, the surrounding particles act as supporting structures 
that convert undulatory motion into net forward motion. Here, the effective propulsion 
depends on the difference between the applied load by the nematode 
undulations and the rigidity of the granular media \cite{shimada, li}. The rigidity, 
or ability to withstand deformation, depends on the local particle area density and becomes 
more solid-like at larger densities, until reaching the jamming limit 
$\phi_J \approx 0.84$ \cite{donev}. This rigidity also depends on the size 
dispersity, which affects the amount of disorder of the granular media 
\cite{muthuswamy, geng2}. Polydisperse samples are more disordered 
and therefore present a more ``rigid'' quasi-static fluidic environment, 
whereas monodisperse samples are less disordered and are more easily sheared, therefore
appear more fluid-like to the \textit{C. elegan} \cite{bardenhagen}.

Finally, an estimate of the propulsive force ($F_{P}$) can be
obtained by computing the work done on the surrounding particles from 
nematode undulations, $W = F_PA - F_Dx$, where $A$ is the nematode 
beating amplitude, $F_D$ is Stokes' drag, and $x$ is the average displacement 
of a group of particles. The kinetic energy of the particles is computed from 
particle tracking data. The propulsive force is estimated by
$F_P= \pi \rho d^3 [N v_g^2/12 A ] + 3\pi\eta d [Nv_g x/A].$
Here $\rho$ is the particle density, $d$ is the particle diameter,
$\eta$ is the buffer dynamic viscosity, $N$ is the average number of
particles moving, and $v_g$ is the average group velocity of the
particles per undulation. This approximation reveals that the force
of the \textit{C. elegans} on the surrounding media is
$F_P \sim \mathcal{O}(10^{-10})$ Newtons. This value is similar to the force
generated by a slender-body swimming in a viscous fluid \cite{Sznitman2} and is
only slightly less than values measured for crawling \textit{C. elegans} 
on substrates using force sensing pillars \cite{doll}.

\section{Conclusions}

In conclusion, we have shown that live organisms can exploit the
material properties of the surrounding complex media to achieve efficient
propulsion. Surprisingly, the addition of particles to a buffer solution
does not hinder the \textit{C. elegans} ability to move. In fact, 
the nematode's propulsion speed is enhanced by the presence of particles
and is not dependent on particle area density. The size dispersity on
the other hand, does affect the locomotion gait of the nematode at 
high area densities.

The adaptive motility behavior of \textit{C. elegans} in
wet granular media may explain the consistent performance seen not
only in nematodes \cite{korta}, but also in flagellated bacteria and
human spermatozoa. Similar mechanisms have been suggested in the
enhanced movement of Leptospira \cite{kaiser} and other flagellated
bacteria in non-Newtonian fluids. Hence, it is possible that other
forms of microscopic propulsion are also tuned to biological fluids
which possess anisotropies. In addition, we are able to study the response and 
rearrangement of the granular media and its dependence on packing fraction
and particle size dispersity free of external forces, such as gravity. This
suggests an interesting method to probe more complex anisotropic
media that could depend on shape or particle interactions. 

\acknowledgments
We thank X. Shen, D. J. Durian, T. Shinbrot, P. B. 
Umbanhowar, and J. P. Gollub for providing helpful comments. This 
work was partially funded by NSF CAREER award CBET-0954084.

\bibliographystyle{eplbib}
\bibliography{nematode}

\end{document}